\documentclass{elsart}

\usepackage{amssymb}
\usepackage{amsmath}
\usepackage{graphicx}
\usepackage{bm}
\usepackage{epsfig}
\begin{document}

\begin{frontmatter}


\title{ Comparison of weighted and unweighted histograms }
\author{N.D. Gagunashvili}
\ead{nikolai@unak.is}
\address {University of Akureyri, Faculty of Information Technology, Borgir, v/Nordursl\'od, IS-600 Akureyri, Iceland }
\begin{abstract}
 Two  modifications  of the  $\chi^2$ test for  comparing 
 usual (unweighted) and weighted histograms  and  two weighted histograms 
are proposed. Numerical examples illustrate an  application of the tests for the
 histograms with  different statistics of events.  Proposed tests can be used 
for the comparison of experimental data histograms against simulated data
 histograms and two simulated data histograms.
\end{abstract}
\end{frontmatter}
\section{Introduction}
A frequently used technique in data analysis is the comparison of histograms. 
First suggested by Pearson \cite{pearson}  the $\chi^2$  test of
 homogeneity   is  used widely  for  comparing usual (unweighted)  histograms.
The modified $\chi^2$  test for comparison of weighted and unweighted
 histograms   recently was  proposed in  \cite{gagunashvili}.

This paper develops the ideas presented in \cite{gagunashvili}. From  this
 development, two new results are presented.
  First, the $\chi^2$ test for comparing  weighted and unweighted histograms
 is improved so that it can be applied for histograms with lower  minimal
 number of events in a bin   than is recommended in  \cite{gagunashvili}.
  And secondly, a   new $\chi^2$ test is proposed  for the  comparison two
 weighted histograms.

The paper is organized as follows. In section 2  the usual  $\chi^2$ test and
 its application for the  comparison of usual unweighted histograms   is
 discussed.
 Tests for the  comparison of  weighted and unweighted histograms and two
 weighted histograms are proposed in sections  3 and 4 respectively. 
 In  section 5 the  tests are  illustrated  and verified  by a numerical
 example and experiments.

\section{ $\chi^2$ test for comparison two (unweighted) histograms} 
 Without limiting the general nature of
 the discussion, we consider two  histograms with the  same
 binning and the  number of bins equal to $r$.
Let us denote the number of events in the $i$th bin in the first histogram as 
$n_{i}$ and as  $m_{i}$ in the second one. The total number of events in the
 first histogram is equal to $N=\sum_{i=1}^{r}{n_i}$,  
and  $M=\sum_{i=1}^{r}{m_i}$  in the second histogram.

The  hypothesis of homogeneity \cite{cramer} is that the two histograms
  represent random  values with  identical distributions.  
  It is equivalent that there  exist  $r$ constants $p_1,...,p_r$,
 such that $\sum_{i=1}^{r} p_i=1$, 
 and the probability  of  belonging  to the  $i$th bin for some  measured value
 in both experiments is  equal to $p_i$.
 The number of events in the $i$th bin is a random variable
 with a distribution  approximated  by a  Poisson probability distribution
 $e^{-Np_i}(Np_i)^{n_i}/n_i!$ for the first histogram and with 
distribution  $e^{-Mp_i}(Mp_i)^{m_i}/m_i!$ for the second histogram. 
 If the hypothesis of homogeneity is valid, then the  maximum likelihood
estimator of  $p_i,i=1,...,r$,  is 
\begin{equation}
 \hat{p}_i= \frac{n_{i}+m_{i}}{N+M} \, \text{ ,} \label{phat}
\end{equation}
and then
\begin{equation} 
X^2=\sum_{i=1}^{r}{\frac{(n_{i}-N\hat{p}_i)^2}{N\hat{p}_i}}
+\sum_{i=1}^{r}{\frac{(m_{i}-M\hat{p}_i)^2}{M\hat{p}_i}} =\frac{1}{MN} \sum_{i=1}^{r}{\frac{(Mn_i-Nm_i)^2}{n_i+m_i}} \label{xsquar}
\end{equation}
has approximately a $\chi^2_{(r-1)}$ distribution \cite{cramer}.

The  comparison   procedure  can include an  analysis of the residuals which
 is often helpful in identifying the bins of histograms responsible for a 
significant overall $X^2$ value. Most convenient for analysis are the 
 adjusted (normalized) residuals \cite{haberman} 
\begin{equation}
r_i=\frac{n_{i}-N\hat{p}_i}{\sqrt{N\hat{p}_i}\sqrt{(1-N/(N+M))(1-(n_i+m_i)/(N+M))}} \, \text{.} \label{resunweighted1}
\end{equation}
 If hypotheses of  homogeneity are valid then 
residuals $r_i$ are approximately independent and identically distributed
 random variables  having  $\mathcal{N}(0,1)$  distribution. Notice that
 residuals (\ref{resunweighted1}) are related with the first histogram and
  residuals related with the second histogram are:
\begin{equation}
r_i'=\frac{m_{i}-M\hat{p}_i}{\sqrt{M\hat{p}_i}\sqrt{(1-M/(N+M))(1-(n_i+m_i)/(N+M))}} \, \text{.} \label{resunweighted2}
\end{equation} 
As  $r_i=-r_i'$,  it makes sense either to  use residuals (\ref{resunweighted1}) or (\ref{resunweighted2}).

The application of the  $\chi^2$ test has restrictions related to the
  value of the expected frequencies $Np_i,Mp_i, i=1,...,r$.   
A conservative rule formulated in \cite{lewontin}  is that all
 the expectations  must be 1 or greater for both histograms. The authors  point out that this
 rule is extremely conservative and in the  majority of cases the $\chi^2$ test may
 be used for histograms  with expectations in excess of 0.5 in the smallest
 bin. In  practical cases when  expected frequencies are not known the estimated expected  frequencies  $M\hat{p}_i$, $N\hat{p}_i, i=1,...,r$ can be used.  
\section{Unweighted and weighted histograms comparison}
A simple  modification of the  ideas described above can be used for the
 comparison of the usual (unweighted) and 
weighted histograms. Let us denote the number of events in the $i$th bin in the unweighted histogram as
$n_{i}$ and  the common weight of events in the $i$th bin of the
weighted histogram as $w_i$. The total number of events in the
 unweighted histogram is equal to $N=\sum_{i=1}^{r}{n_i}$ and  the total
 weight of events in the weighted histogram is equal
 to $W=\sum_{i=1}^{r}{w_i}$.

 Let us formulate the hypothesis of identity of an unweighted histogram 
to a weighted histogram so that  there  exist  $r$ constants $p_1,...,p_r$,
 such that $\sum_{i=1}^{r} p_i=1$, and the probability  of  belonging  to the  $i$th bin for some  measured value
  is  equal to $p_i$ for the  unweighted histogram and expectation value of weight $w_i$ equal to $Wp_i$ for the  weighted histogram.
The number of events in the $i$th bin is a random
variable  with distribution  approximated  by the  Poisson probability distribution
  $e^{-Np_i}(Np_i)^{n_i}/n_i!$ for the  unweighted  histogram.
The weight $w_i$ is a random variable with a distribution approximated  by 
 the normal probability  distribution  $ \mathcal{N}(Wp_i,\sigma_i^2)$, where $\sigma_i^2$ is the  variance of the  weight  $w_i$.  
 If we replace the variance  $\sigma_i^2$ with estimate  $s_i^2$ (sum of squares of weights of events in the $i$th bin) and 
 the hypothesis of identity is valid, then the   maximum likelihood
estimator of  $p_i,i=1,...,r$,  is
\begin{equation}
 \hat{p}_i= \frac{Ww_i-Ns_i^2+\sqrt{(Ww_i-Ns_i^2)^2+4W^2s_i^2n_i}}{2W^2} \, \text{.} \label{phat2}
\end{equation}
We may then use the test statistic
\begin{equation} 
X^2=\sum_{i=1}^{r}{\frac{(n_{i}-N\hat{p}_i)^2}{N\hat{p}_i}}
+\sum_{i=1}^{r}{\frac{(w_{i}-W\hat{p}_i)^2}{s_i^2}}   \label{xsquarweigh}
\end{equation}
and it is plausible that this  has approximately a $\chi^2_{(r-1)}$ distribution.

This  test, as well as the  original one \cite{cramer}, has a restriction
 on the expected frequencies. The expected frequencies 
  recommended for the  weighted histogram  is  more than 25.
The value of the  minimal expected frequency can be decreased down to 10 for
 the case when the weights of the events are close to constant.
In the case of a weighted histogram if the number of events is unknown, then we can apply this recommendation for the equivalent number of events as $n_i^{equiv}={w_i^2}/{s_i^2} \, \text{.}$ The minimal   expected frequency for an  unweighted histogram must be 1. 
Notice that any usual (unweighted)  histogram can be considered as a weighted histogram with events that have constant weights equal to 1.

The  variance $z_i^2$  of the difference between the weight $ w_{i}$ and the estimated expectation value of the weight is  approximately  equal to: 
\begin{equation}
\begin{split} 
z_i^2=Var(w_{i}-W\hat{p}_i)=N\hat{p}_i(1-N\hat{p}_i)\biggl(\frac{Ws_i^2}
{\sqrt{(Ns_i^2-w_iW)^2+4W^2s_i^2n_i}}\biggr)^2\\
+\frac{s_i^2}{4}\biggl(1+\frac{Ns_i^2-w_iW}
{\sqrt{(Ns_i^2-w_iW)^2+4W^2s_i^2n_i}}\biggr)^2 \, \text{.} \qquad \qquad \qquad
\end{split}
\end{equation}
The  residuals 
\begin{equation}
r_i=\frac{w_{i}-W\hat{p}_i}{z_i} \label{rere}
\end{equation}
have approximately a normal distribution with mean equal to 0 and
 standard deviation  equal to 1.
\section{Two weighted histograms comparison}
Let us denote the  common  weight of events of the $i$th bin in the first histogram as
$w_{1i}$ and as  $w_{2i}$ in the second one. The total  weight of events in the
 first histogram is equal to $W_1=\sum_{i=1}^{r}{w_{1i}}$,
and  $W_2=\sum_{i=1}^{r}{w_{2i}}$  in the second histogram.

 Let us formulate the hypothesis of
 identity of   weighted histograms  so that  there  exist  $r$ constants $p_1,...,p_r$,
 such that $\sum_{i=1}^{r} p_i=1$, and  also  expectation value of weight $w_{1i}$ equal to $W_1p_i$ and
expectation value of weight $w_{2i}$ equal to $W_2p_i$.
Weights in both the histograms are random variables with  distributions which
 can be
 approximated by a normal probability distribution  $\mathcal{N}(W_1p_i,\sigma_{1i}^2)$
 for the first histogram and by a distribution  $\mathcal{N}(W_2p_i,\sigma_{2i}^2)$ for the
 second.  Here  $\sigma_{1i}^2$ and  
 $\sigma_{2i}^2$ are the  variances of $w_{1i}$ and $w_{2i}$ with estimators $s_{1i}^2$
 and $s_{2i}^2$ respectively. If the hypothesis of identity is valid,
 then  the  maximum likelihood  and Least  Square Method  estimator 
 of  $p_i,i=1,...,r$,  is
\begin{equation}
\hat{p}_i=\frac{w_{1i}W_1/s_{1i}^2+w_{2i}W_2 /s_{2i}^2}{W_1^2/s_{1i}^2+W_2^2/s_{2i}^2} \, \text{.}  \label{plms}
\end{equation}
We may then use the test statistic
\begin{equation}
X^2=\sum_{i=1}^{r}{\frac{(w_{1i}-W_1\hat{p}_i)^2}{s_{1i}^2}}
+\sum_{i=1}^{r}{\frac{(w_{2i}-W_2\hat{p}_i)^2}{s_{2i}^2}}=\sum _{i=1}^{r}{\frac{(W_1w_{2i}-W_2w_{1i})^2}{W_1^2s_{2i}^2+W_2^2s_{1i}^2}}   \label{xsquarweigh}
\end{equation}
and it is plausible that this  has approximately a $\chi^2_{(r-1)}$ distribution. The normalized or studentised residuals \cite{seber}
\begin{equation}
  r_i=\frac{w_{1i}-W_1\hat{p}_i}{s_{1i}\sqrt{1-1/(1+W_2^2s_{1i}^2/W_1^2s_{2i}^2)}}
\end{equation}
have approximately a normal distribution with mean equal to 0 and
 standard deviation 1. A recommended minimal expected frequency is  equal to 10 for the proposed test.

\section{Numerical example and experiments}
The method described herein is now  illustrated with an example.
We take a  distribution
\begin{equation}
\phi(x)=\frac{2}{(x-10)^2+1}+\frac{1}{(x-14)^2+1} \label{weight}
\end{equation}  
 defined on the interval $[4,16]$.  Events distributed 
according to the formula (\ref{weight}) are simulated  to create the unweighted
 histogram.
 Uniformly  distributed events are simulated for the  weighted histogram 
 with  weights calculated by  formula (\ref{weight}).
 Each histogram has the same  number of bins: 20.    
 Fig. 1 shows the result of comparison of the  unweighted histogram with
200 events  (minimal expected frequency equal to one) and the weighted histogram with 500 events (minimal expected frequency equal to 25)
\begin{figure}
\centering
\includegraphics[width=15cm]{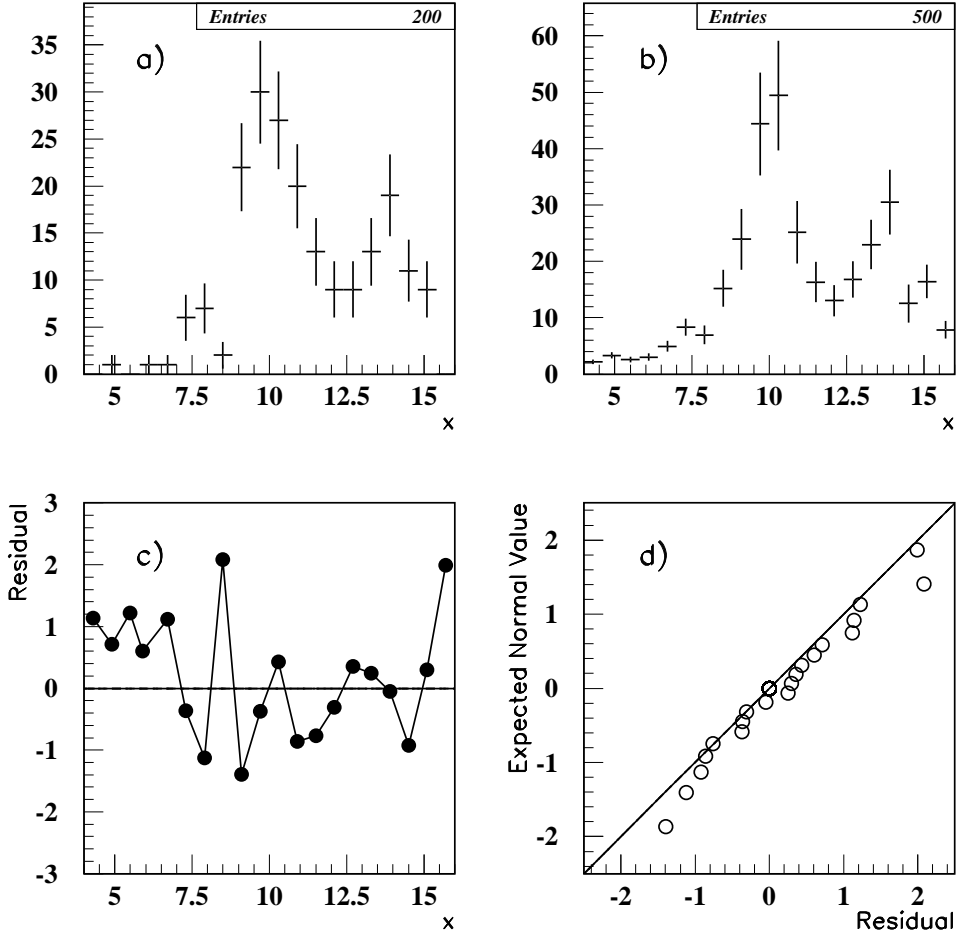}
\vspace*{-1.7 cm}
\caption {An example of comparison of the unweighted histogram with 200 events
 and the  weighted histogram with 500 events: a) unweighted histogram;
 b) weighted
 histogram; c) normalized residuals plot; d) normal Q-Q plot of residuals. }
\end{figure}

 The value of the test statistic
$X^2$ is equal to 21.09 with $p$-value equal to 0.33, therefore the
 hypothesis
 of  identity of the two histograms  can be accepted. The behavior of the
 normalized residuals  plot (see Fig. 1c) and the  normal Q-Q plot (see Fig. 1d) of  residuals  are
 regular and we cannot identify the  outliers or bins with a big influence on
 $X^2$.

  To investigate the dependence of the distribution of  the test statistics
 from the  number of events all three tests  were considered.

{\em The  comparison of pairs of  unweighted  histograms with 
different minimal expected frequencies  was considered 
(Pearson's chi square test).}  
 Unweighted histograms with  minimal  expected frequencies equal to one
 (200 events), 2.5 (500 events) and 5  (1000 events) where  simulated.  Fig. 2 shows the    Q-Q plots  of  $X^2$ statistics for different  pairs of  histograms. In each case 10000 pairs of histograms were simulated. 
\begin{figure}
\centering
\includegraphics[width=15cm]{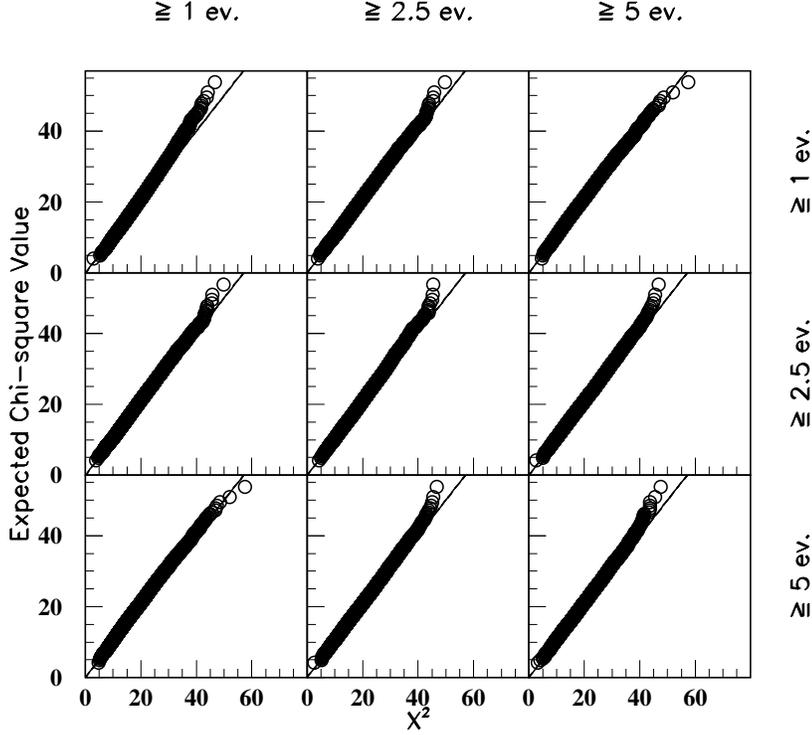}
\vspace*{-2.8 cm}
\caption { Chi-square  Q-Q plots  of $X^2$ statistics for two unweighted histograms  with different  minimal expected frequencies. }
\end{figure}
 As we can see for all cases the real distributions of test statistics  are close to the  theoretical $\chi^2_{19}$  distribution.

{\em The  comparison of pairs of  unweighted  and weighted histograms with 
different minimal expected frequencies
  was considered  using the test proposed in section 3 above.} 
Unweighted histograms with  minimal  expected frequencies equal to one
 (200 events), 2.5  (500 events) and 5  (1000 events) where simulated. Furthermore
  weighted histograms with minimal expected frequencies equal to  10
 (200 events), 25 (500 events) and 50  (1000 events) where simulated.
  Fig. 3 shows the    Q-Q plots  of  $X^2$ statistics for different  pairs of  histograms.
\begin{figure}
\centering
\includegraphics[width=15cm]{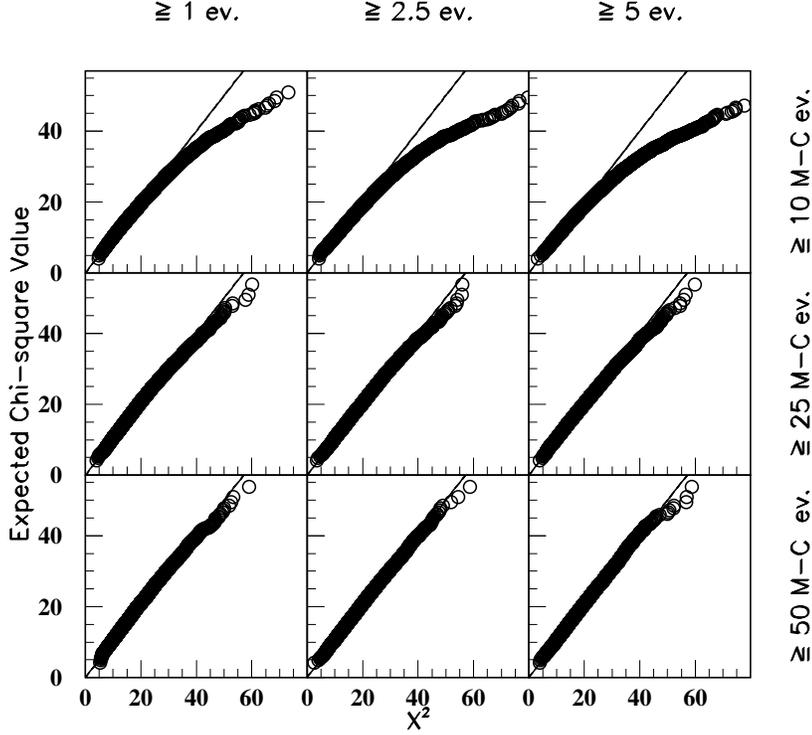}
\vspace*{-2.8 cm}
\caption { Chi-square  Q-Q plots  of $X^2$ statistics for  unweighted and weighted histograms  with different  minimal expected frequencies.}
\end{figure}
 As we can see the  real  distribution of test statistics obtained for
 minimal expected frequency  of  weighted events, equal to 10,  has a  heavier tail than the  theoretical  $\chi^2_{19}$
 distribution. This  means that the  $p$-value calculated with the   theoretical
 $\chi^2_{19}$  distribution is lower than the real $p$-value and any decision
 about the rejection of the  hypothesis of identity of the two  distributions is
 conservative. The distributions of test statistics for the minimal expected frequencies  25  and 50  are  close to the theoretical distribution.
This  confirms that the minimal  expected  frequency  25 is reasonable restriction for the  weighted  histogram  for this test.

{\em The  comparison  of  two  weighted histograms with 
different minimal expected frequencies  was considered using the test proposed  in section 4 above.}  
 Weighted histograms with minimal expected frequencies equal to
 10 (200 events), 25 (500 events) and 50 (1000 events) where simulated.
  Fig. 4 shows the    Q-Q plots  of  $X^2$ statistics for different  pairs of
  histograms.
\begin{figure}
\centering
\includegraphics[width=15cm]{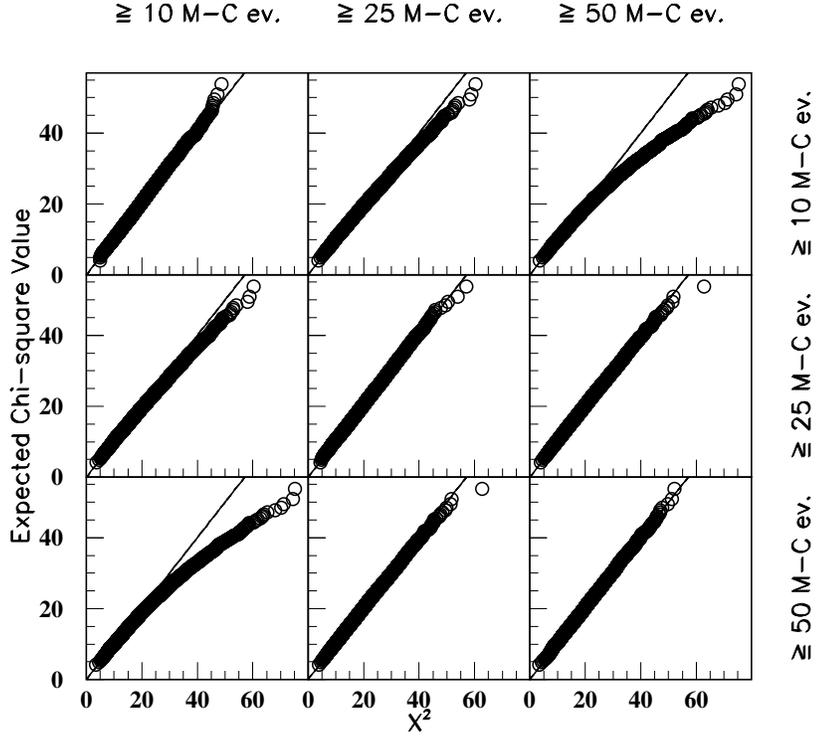}
\vspace*{-2.8 cm}
\caption { Chi-square  Q-Q plots  of $X^2$ statistics for two weighted histograms  with different  minimal expected frequencies.}
\end{figure}
 As we can see the real distributions of the test statistics  are close to the
 theoretical $\chi^2_{19}$  distribution if the  minimal expectations of the
 two histograms  are close to each other, it is  in all cases excluding case
 (10, 50). For the case when the difference in expectations are big (10, 50) the real distribution of the test statistics   has a  heavier tail than the  theoretical  $\chi^2_{19}$. 

To verify the proposed tests  two further numerical experiments were performed.

For the first case  unweighted histograms with minimal expected frequencies equal to  10 (2000 events), 25 (5000 events) and 50  (10000 events) were simulated. These histograms were compared to an unweighted histogram with 10 or more expected frequencies by the three  methods  described above.
 Fig. 5 shows the    Q-Q plots  of  $X^2$ statistics for different  pairs of  histograms.
\begin{figure}
\centering
\includegraphics[width=15cm]{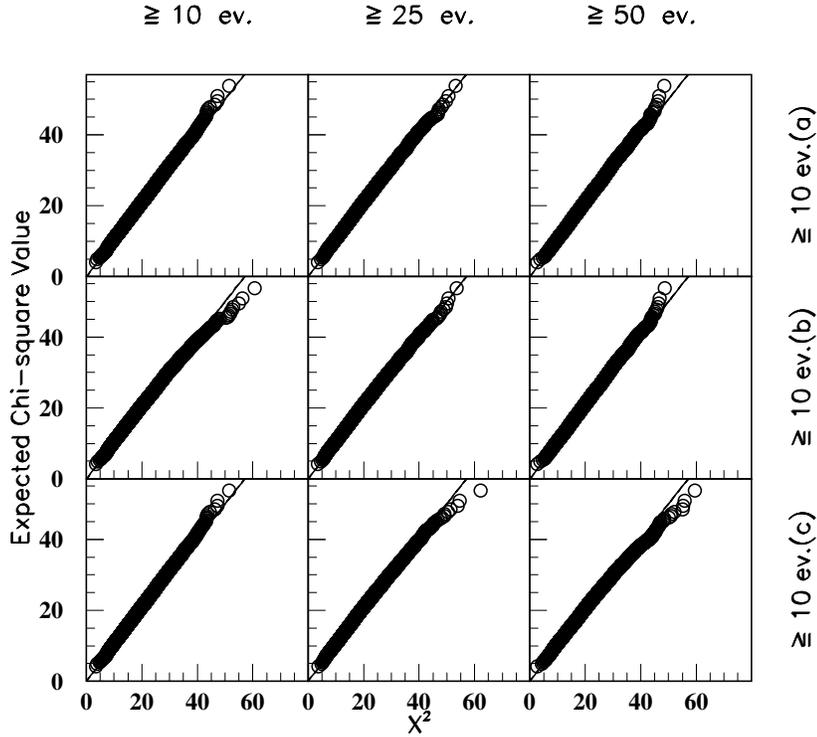}
\vspace*{-2.8 cm}
\caption { Chi-square  Q-Q plots  of $X^2$ statistics for two unweighted histograms  with different  tests: a) Pearson's chi square  test; b) proposed in this article test for unweighted and weighted histograms; c) proposed in this article test for two weighted histograms.}  
\end{figure}
 As we can see  the real distributions of the test statistics  are close to the theoretical $\chi^2_{19}$  distribution for all three tests.

For the second  case  unweighted histograms with  minimal expected frequencies  equal to one (200 events), 2.5 (500 events) and 5  (1000 events) were simulated. These histograms were compared to an  unweighted histogram with
10 or more expected frequencies by the first two  methods  described above.
 Fig. 6 shows the    Q-Q plots  of the  $X^2$ statistics for different
  pairs of  histograms.
\begin{figure}
\centering
\includegraphics[width=15cm]{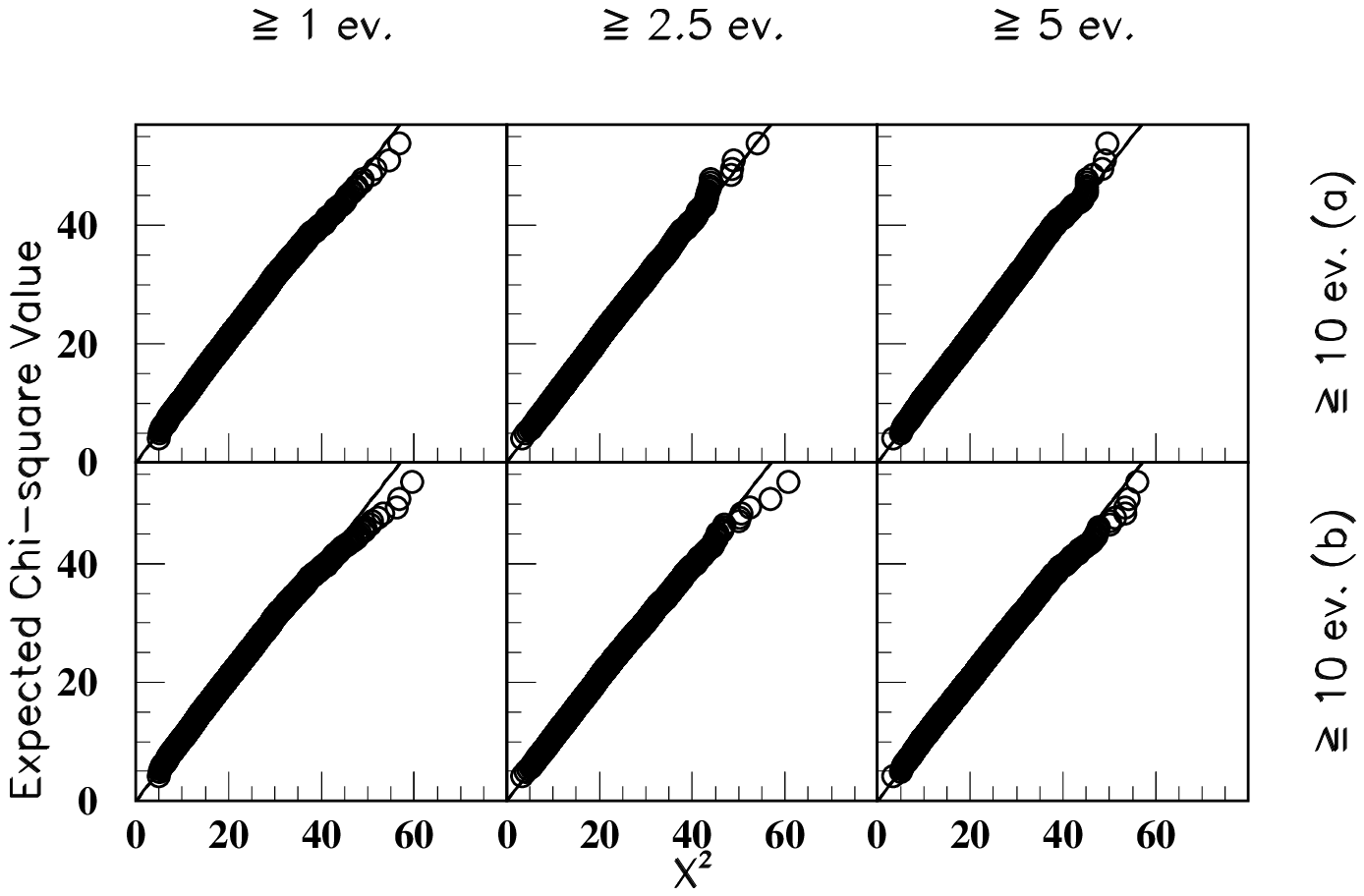}
\vspace*{-5. cm}
\caption { Chi-square  Q-Q plots  of $X^2$ statistics for two unweighted histograms  with different  tests: a) Pearson's chi square test; b) proposed in this article test for unweighted and weighted histograms. }
\end{figure}
 As we can see for all cases the real distributions of the  test statistics
  are close to the theoretical $\chi^2_{19}$  distribution.
 Also the  real distributions of the test statistics for the proposed method
 of comparison of  unweighted and weighted histograms (see Fig. 6b) do not
 have heavy tails as is the case for a  weighted histogram with weights calculated according formula (\ref{weight}) (see Fig. 3).
This example confirms that the minimal expected frequency equal to  10 
is enough for the application of the   method of comparison of unweighted and weighted histograms  if the weights  of the events are close to a constant for the  weighted histogram.
\section{Conclusions}
A chi square test for comparing the  usual (unweighted) histogram and the weighted
 histogram, together with a test for comparing  two weighted histograms were proposed. 
In both cases formulas for normalized residuals were presented that can be useful for the identifications of bins that are outliers, or bins that have a big influence on $X^2$.
For the first test the recommended minimal expected frequency of events is equal
 to 1 for an unweighted histogram and 10-25 for a weighted histogram.
For the second test the recommended minimal expected frequency is  equal to  10.
Numerical examples illustrated an application of  the method for the
 histograms with different statistics of events and confirm that the proposed
 restrictions related with the expectations are reasonable. 
The  proposed in this paper approach can be  generalized  for a comparison of  several
 unweighted and weighted histograms or just weighted histograms.
 The test statistic  has approximately a  $\chi^2_{(r-1)(s-1)}$ distribution
 for $s$  histograms with $r$ bins. 
\ack
The author is very grateful to Steffen Lauritzen (University of Oxford) who
 suggested  idea of improving the method for comparing  weighted and
 unweighted histograms, and to Mark O'Brien (University of Akureyri) for
 reading the paper in draft form  and making constructive comments.   


\begin{thebibliography}{9}
\bibitem{pearson}
Pearson, K., 1904. On the Theory of Contingency and Its Relation to Association
 and Normal Correlation. Drapers' Co. Memoirs, Biometric Series No. 1, London. 
\bibitem{gagunashvili}
Gagunashvili, N., 2006.  $\chi^2$ test for comparison of weighted and
 unweighted histograms.  Proceedings of Conference "Oxford 2005, Statistical
 Problems in Particle Physics, Astrophysics and Cosmology". Imperial College
 Press, London, 191-192. 
\bibitem{cramer}
Cramer, H., 1946. Mathematical methods of statistics. Princeton University Press, Princeton. 
\bibitem{haberman}
Haberman, S.J., 1973. The analysis of residuals in cross-classified tables. Biometrics 29, 205-220.
\bibitem{lewontin}
Lewontin, R.C. and  Felsenstein, J., 1965.  The robustness of homogeneity test
in $2 \times N$ tables. Biometrics 21, 19-33.
\bibitem{seber}
Seber,  G.A.F., Lee,  A.J., 2003,  Linear Regression Analysis. John Wiley \& Sons Inc., New York.

\end{thebibliography}
\end{document}